\documentclass[12pt]{iopart}

\usepackage{amssymb}
\usepackage{epsfig}
\usepackage{cite}

\newcommand{\igr}[2][]{\includegraphics[#1]{#2}}

\begin{document}

\title[Intraband electron focusing in bilayer graphene]{Intraband electron focusing in bilayer graphene}

\author{Csaba G. P\'eterfalvi$^1$, L\'aszl\'o Oroszl\'any$^2$,
Colin J. Lambert$^1$ and J\'ozsef Cserti$^2$}
\address{$ˆ1$ Department of Physics, Lancaster University, LA1 4YB, UK}
\address{$ˆ2$ Department of Physics of Complex Systems, E\"otv\"os University, H-1117
Budapest, P\'azm\'any P\'eter s\'et\'any 1/A, Hungary}
\ead{c.peterfalvi@lancaster.ac.uk}

\begin{abstract}
We propose an implementation of a valley selective electronic Veselago
lens in bilayer graphene. We demonstrate that in the presence of an
appropriately-oriented potential step, low-energy electrons radiating
from a point source can be re-focused coherently within the same band.
The phenomena is due to the trigonal warping of the band structure
that leads to a negative refraction index. We show that the interference
pattern can be controlled by an external mechanical strain.
\end{abstract}

\submitto{\NJP}
\maketitle

\section{Introduction}

Since its first experimental isolation~\cite{Novoselov2005,Novoselov2006} graphene
has had a profound impact on the solid state community. The past year
has seen reinvigorated experimental and theoretical investigations
of bilayer graphene, as sufficiently-clean samples have been fabricated
to probe the low-energy spectrum~\cite{Mayorov2011}. A key conclusion
of these investigations is that mechanical distortion as well as electron-electron interaction
can lead to profound changes in the topology of the band structure
and to symmetry breaking in the electronic system~\cite{Lemonik2010,Mucha-Kruczynski2011,Gradinar2012}.
In recent years, new ideas in electron optics have utilized the exotic
electronic structure of graphene and other novel materials such as
topological insulators. Electron lenses utilizing negative refraction~\cite{Cheianov2007,Cserti2007,Hassler2010,Park2011}
and valley-polarized electron beam splitters~\cite{Garcia-Pomar2008,pereira_jr_valley_2009,abergel_generation_2009,Wang2010}
among others have been proposed to manipulate electron beams in mesoscopic
devices. Apart from the manipulation of the valley degree of freedom,
graphene devices have already been proposed for spin polarized electron beam engineering
using magnets~\cite{michetti_electric_2010}, and for spatial separation of electrons and holes
using normal-superconducting-normal junctions~\cite{gomez_selective_2012}.

Unlike 2DEGS (two-dimensional electron gas) in more conventional semiconductors~\cite{vanHouten1988,Topinka2003},
which require a magnetic field to provide the focussing of electrons
and imaging of the interference pattern of injected electrons is hindered
by the fact that the 2DEG is often buried inside a heterostructure~\cite{Gilbertson2011},
graphene-based devices can be self-focussing
and their interference patterns are exposed on the surface. In the
present manuscript we propose an electron-optical device, whose valley-dependent
interference pattern can be controlled by strain. By analogy with conventional 2DEGS,
where the control of focussing by an external magnetic field leads
to sensitive magnetic field sensors even at room temperature~\cite{Aidala2007},
our graphene-based device forms a basis for a sensitive detector of
strain.

\begin{figure}
\igr[bb=0bp 0bp 1269bp 732bp,clip,width=1\columnwidth]{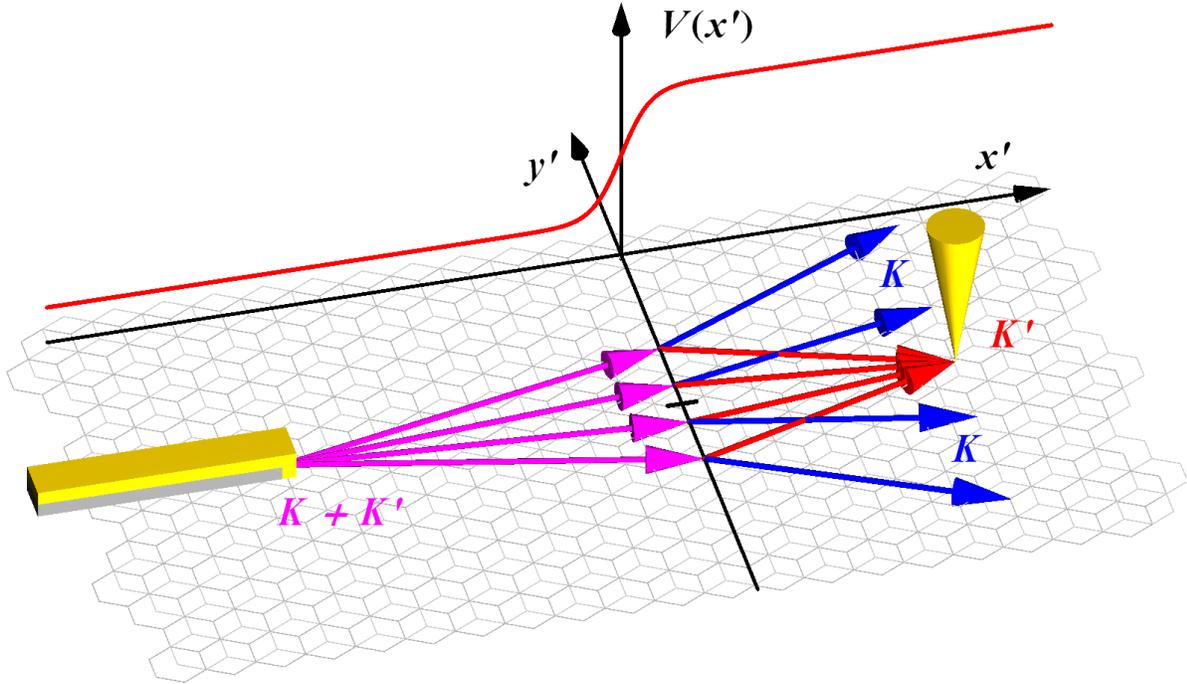}\caption{\label{fig:Setup}Single band focusing in bilayer graphene with a
potential step. A particle emitted from a source disperses or focuses
depending on which valley it came from. Measuring the maxima of the
transmitted electron density with an STM tip can be used to indirectly
gain information about the low energy topology of the band structure.}
\end{figure}

\section{Electron Focusing in Novel Materials}

Focusing of electrons in graphene by a planar $pn$ junction has first been
proposed by Cheianov and coworkers~\cite{Cheianov2007}. In their
proposal, particles are focused due to the fact that in the valence
band, the group velocity points in the opposite direction to the wavevector.
Hence electrons emitted from a point source on one side of a $pn$
junction converge on the other side at a focal point, thereby realizing
a Veselago lens~\cite{Veselago1968,Pendry2000}. Circular $pn$ junctions
have also been investigated recently in single layer~\cite{Cserti2007}
and bilayer~\cite{Peterfalvi2009} graphene, where apart from the focal
point, a hierarchy of caustics was discovered and described using
the semiclassical approach of catastrophe optics~\cite{Peterfalvi2010}.
Later it was pointed out that warping of the band structure~\cite{McCann2006}
can also induce a focusing effect even when particles on the two sides
of the junction are in the same band~\cite{Hassler2010}. This was
demonstrated with the hexagonally warped surface states of the three
dimensional topological insulator $\mbox{Bi}_{2}\mbox{Te}_{3}$~\cite{Fu2009}.
In this work we propose a flat interface setup (shown in Fig.~\ref{fig:Setup}) that can focus electrons
in bilayer graphene due to the trigonal warping~\cite{McCann2006}
of the band structure. In this section we review some aspects of electron
optics with regards to focusing particles by a flat interface.

For a particle incident on the barrier with wavevector $\mathbf{k}_{i}$
and group velocity $\mathbf{v}_{i}=\left.\left(\partial_{\mathbf{k}}E(\mathbf{k})/\hbar\right)\right|_{\mathbf{k}=\mathbf{k}_{i}}$,
the refractive index $n$ at the barrier is defined by the states
that conserve the momentum parallel to the interface and propagate
away from the interface. Let $\mathbf{k}_{t/r}$ and $\mathbf{v}_{t/r}$
denote the wavevector and group velocity of the transmitted/reflected
particles respectively. If the velocities $\mathbf{v}_{i/t}$ make
an angle $\phi_{i/t}$ with the normal of the barrier then the refractive
index is
\begin{equation}
n=\frac{\sin(\phi_{i})}{\sin(\phi_{t})}.
\end{equation}
In general, $n$ depends on the direction of $\mathbf{k}_{i}$.

\begin{figure*}
\igr[bb=100bp 0bp 1168bp 277bp,clip,width=1\textwidth]{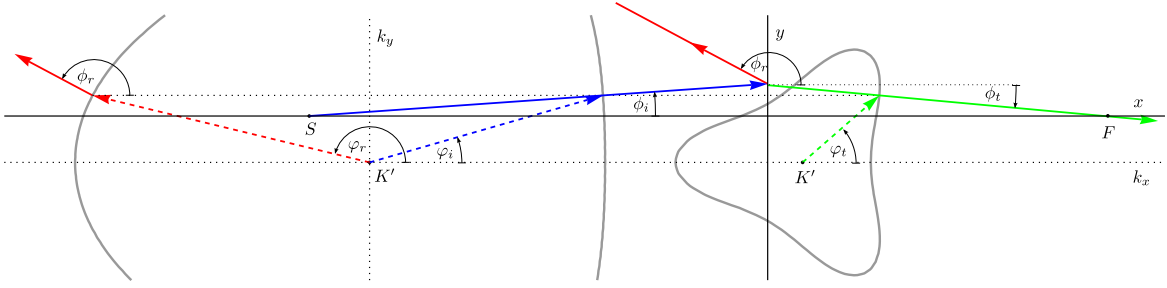}\caption{\label{fig:setup_real_and_k_space-single-ray}
Real space trajectory and $k$-space directions of a single particle incident on a potential
step in the valley $K$' in a bilayer graphene junction at $w=0$
and $\alpha=0$, \emph{i.e.} where the $(x,y)$ and $(x',y')$ systems
coincide. Angles $\varphi$ denote the direction of the wavevectors,
while angles $\phi$ stand for the direction of the group velocities.
Note that in this anisotropic system the group velocity and the wavevector
of a given state are in general not parallel. A particle emitted from
the point source $S$ at $-2.07\mu\mbox{m}$ on the $x$-axis, hits
the boundary at the $y$-axis with an angle of incidence $\phi_{i}$
and along with electrons on neighbouring trajectories, gets diverted
towards a cusp type caustic, the maximum of which we call the focus
F, for the sake of simplicity. The electrons' energy is $80\mbox{ meV}$
on the left and $10\mbox{ meV}$ on the right side of the junction.
Blue arrows represent the incident particle, green arrows the transmitted,
and red arrows the reflected ones. The wavevectors are denoted by
dashed arrows and the classical trajectory of the particle by solid
lines. Solid lines depict the electrons' dispersion curves on both sides of the
junction. Note that the curvature has different sign on the two sides
for particles close to the optical axis.}
\end{figure*}

The intraband focusing effect of a planar interface
is closely related to the concave geometry of the $E(\mathbf{k})$-energy
contours of the dispersion relation. It can be shown that electrons
radiating from a point source into a small region in $\mathbf{k}$-space
will be focused on the other side of the junction if and only if the
curvatures of the energy contours on the two sides at the corresponding
\textbf{$\mathbf{k}$}-wavevectors
differ in sign. If this holds around the $\mathbf{k}$-vectors aligned
with the optical axis, that is for rays that are nearly perpendicular
to the interface, then the different signs of the curvatures result
in a negative refractive index, which is however not a necessary condition
for focusing if we move away from the optical axis. In particular,
if the $E(\mathbf{k})$-contours are locally symmetric to the direction
of the optical axis, then a simple, direct relation can be formulated
between the refractive index for these rays and the curvatures on
the two sides~\cite{Hassler2010}:
\begin{equation}
n=\frac{c(\mathbf{k}_{i},E_{i})}{c(\mathbf{k}_{t},E_{t})},
\end{equation}
where $E_{i}$ is the chemical potential on the incoming side of
the junction and $E_{t}$ is the chemical potential on the other side
of the junction. It also follows
that around the optical axis, the negative sign of $n$ is needed
for the focusing phenomenon. In Fig.~\ref{fig:setup_real_and_k_space-single-ray} we show the
trajectory of a single particle hitting a planar potential step in
a bilayer graphene sample. (The barrier coincides with the $y'$-axis.)
A classical trajectory can be parametrized by $\varphi_{i}$ which
determines all the remaining angles.

If a point source is situated at $(-x_{s},0)$, then on the other
side of the interface, a generic trajectory has the form
\begin{equation}
y=x_{s}\tan\left(\phi_{i}\left(\varphi_{i}\right)\right)+x\tan\left(\phi_{t}\left(\varphi_{i}\right)\right).\label{eq:single-ray}
\end{equation}
 Some of the refracted rays touch each other at certain points. The
sets of such points are called caustics, which are envelopes of a
family of rays at which the density of rays is singular. In the classical
theory of geometrical optics, this singularity is unphysical, however,
the results of the semiclassical catastrophe optics~\cite{Berry1980}
agrees well with the quantum mechanical treatment, which predicts
a finite particle density with local maxima~\cite{Peterfalvi2010}.
The focal point of particles corresponds to the global maximum of
a cusp type caustic, which evolves into two fold type caustics. The
coordinates of these curves can be obtained by differential geometry:
\begin{equation}
\partial_{\varphi_{i}}y=0.\label{eq:caustics}
\end{equation}

In order to describe the system quantum mechanically,
one has to solve the Schr\"odinger equation on both sides of the junction
and match the solutions at the boundary. The boundary conditions can
be derived from the Schr\"odinger equation by applying it for an infinitesimally
narrow stripe containing the junction. In the case of the model Hamiltonian
used in this work, these conditions require the wavefunction to be
smooth and the same for its first derivative with respect to the direction
perpendicular to the junction. This can be satisfied with 2 planewave-like
solutions of the homogeneous Schr\"odinger equation on either side of
the junction. For a given $\mathbf{k}_{y'}$, which must be conserved
during the scattering process, there always exist 1 or 2 propagating
waves and 1 or 0 decaying wave, giving always exactly 2 solutions on either sides
needed to satisfy the boundary conditions.
By solving this system of equations, we obtain the coefficients
for the reflected and refracted rays matched to the incident ray.
Since all the calculations can be extended in a straightforward manner
for several rays, for the sake of simplicity, wherever applicable,
we will assume that we only have one refracted wave propagating to
the right direction. To calculate the density of particles,
the contributions of all the refracted rays with
the corresponding transmission amplitudes were summed up in the far
field approximation. By the summation of the rays, a homogeneous distribution
of the source electrons in $\varphi_{i}$ is assumed.

\section{Focusing in Bilayer Graphene }

The central idea of our work is to use bilayer graphene with a potential
step for focusing electrons, where particles remain in the same band.
As we will demonstrate, this idea retains all the neat features of
earlier setups~\cite{Cheianov2007,Hassler2010}, including focusing
and high transmission, and it even incorporates the idea of valley-dependent
electron beam manipulation~\cite{Garcia-Pomar2008,pereira_jr_valley_2009,Fu2009}.

We start with the Hamiltonian of the system under consideration

\begin{equation}
H=\left(\begin{array}{cc}
V(r) & -\frac{1}{2m}p_{-}^{2}+\xi v_{3}p_{+}+w\\
-\frac{1}{2m}p_{+}^{2}+\xi v_{3}p_{-}+w^{*} & V(r) \end{array}\right),\label{eq:zeHamiltonian}
\end{equation}
where $p_{\pm}=p_{x}\pm p_{y}$ are the momentum operators and $\xi=1$ for valley $K$ and
$\xi=-1$ for $K'$. The material parameters of bilayer graphene $m\approx0.035 m_{e}$
and $v_{3}\approx10^{5}\mbox{m/s}$ are defined among others in Ref.~\cite{Koshino2009}.
For very low energies \mbox{($\lesssim 1$ meV)}, the complex parameter $w$ describes symmetry
breaking due to the electron-electron interaction~\cite{Lemonik2010}, but it also describes
the external mechanical strain~\cite{Mucha-Kruczynski2011} for higher energies as well.
The relation between $w$ and the mechanical distortion of the lattice can be formulated as
\begin{equation}
w=\eta(\delta-\delta')\exp(-2i\theta),\label{eq:wmech}
\end{equation}
where $\eta$ is a real constant describing mechanical properties of the lattice,
$\delta > \delta'$ are the principal values of the strain tensor
\mbox{$\frac12(\partial_i d_j + \partial_j d_i)$}, and $\theta$ is
the angle between the principal axes and coordinate axes $(x,y)$.
In the strain tensor, $i$ and $j$ are the $(x,y)$ coordinate-indexes
and $(d_x,d_y)$ denote displacements from the equilibrium positions.
In the expression~(\ref{eq:wmech}) we neglect the interlayer shear shift.
The potential $V(r)$ describes a potential step that is sufficiently smooth on the atomic scale so
that it does not introduce inter-valley scattering, and we can also neglect
a term proportional to the gradient of the potential that arises in
the Schrieffer-Wolff transformation in the derivation of~(\ref{eq:zeHamiltonian})~\cite{McCann2006,Poole2010}.
We assume $V(r)$ to have the form
\begin{equation}
V(\mathbf{r})=V_{1}+V_{2}\Theta(x-\tan(\alpha)y),
\end{equation}
 \emph{i.e.} the potential on the two sides of the step is constant
$V_{1}$ and $V_{2}$ and the barrier makes an angle $\alpha$ with
the crystallographic $\Gamma K$ direction ($k_{x}$-axis). It is
convenient to introduce a coordinate system $(x',y')$ where $y'$
is defined by the planar potential step, and the source is sitting
on the axis $x'$, which means that $x$ and $x'$ and also $y$
and $y'$ make an angle denoted by $\alpha$. We do not take into
account a gap term in the Hamiltonian that accounts for the electrostatic
asymmetry of the two graphene layers, since this would not alter the
main message of the present work. We also neglect second nearest neighbour
hopping in the individual graphene planes that has negligible effect
on the results presented here. In the absence of the potential $V(\mathbf{r})$
the spectrum of~(\ref{eq:zeHamiltonian}) has been studied in~\cite{McCann2006,Mucha-Kruczynski2011}.
For low energies, depending on the value of $w$, the spectrum consists
of four or two massless Dirac cones, whereas at higher energies, the
band structure merges to form only a single pocket per valley. A
typical energy scale of this model is the Lifshitz energy $E_{L}=mv_{3}^{2}/2\approx1\mbox{ meV}$,
which defines the Lifshitz transition of the band structure's topology
from multiple pockets to a single pocket in the absence of strain ($w=0$).
Since the Lifshitz energy is quite small, it was difficult to gain
experimental data from energies below it until recently~\cite{Mayorov2011}.
\begin{figure}
\igr[width=1\columnwidth]{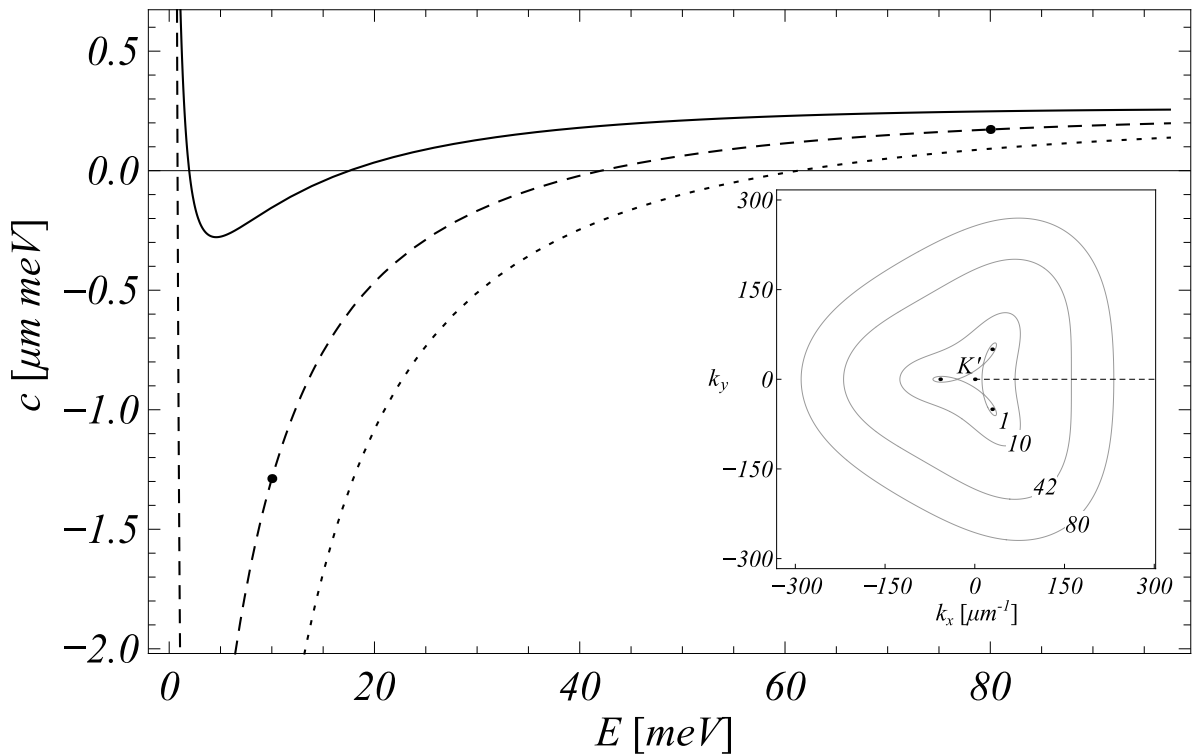}\caption{\label{fig:Curvatures}
Curvature $c$ as the function of energy along the direction indicated by the dashed line in the inset.
Inset shows the energy contours around the $K'$ point for $w=0\mbox{ meV}$.
In the main figure, solid line denotes curvature for $w=-6\mbox{ meV}$,
dashed line is for $w=0\mbox{ meV}$ and dotted line is for $w=6\mbox{ meV}$.
Black dots denote the particles' energy on the left ($80\mbox{ meV}$) and
right ($10\mbox{ meV}$) side of the junction as in Fig.~\ref{fig:setup_real_and_k_space-single-ray}.
The electrons' energy is chosen to find optimal conditions
for focusing, that is a negative and slowly varying refractive index $n$
in a possibly wide range of the angle of incidence $\phi$.}
\end{figure}

As shown in the previous section, focusing is controlled by the curvature
of the band structure. If on the two sides of the potential step there
exist regions in the $k$-space on the given energies with overlapping
projections to the dimension parallel with the junction, and for these
corresponding $k$-regions the curvature of the dispersion relation
has different sign, then the system can focus electrons. In Fig.~\ref{fig:Curvatures}
we present the curvature $c$ of the $E(\mathbf{k})$ dispersion curve
of the considered model as the function of energy along the $\Gamma K$
direction for different values of $w$. It is clear that the curvature
in the absence of $w$ changes sign roughly at $E\approx42.1\mbox{ meV}$.
This energy scale is still sufficiently small compared to the direct
inter layer hopping~\cite{McCann2006,Koshino2009} $\gamma_{1}\approx0.4\mbox{ eV}$,
that is it falls in the energy range where the two band model~(\ref{eq:zeHamiltonian})
is still valid. Since a changing $w$ can produce a large change
in the energy of zero curvature, one can study the focused electrons in bilayer
graphene samples with a planar potential step, and yield information
about the value of $w$, even in samples that are only moderately
gated \mbox{($\sim50$ meV)}.
\begin{figure}
\igr[width=1\columnwidth]{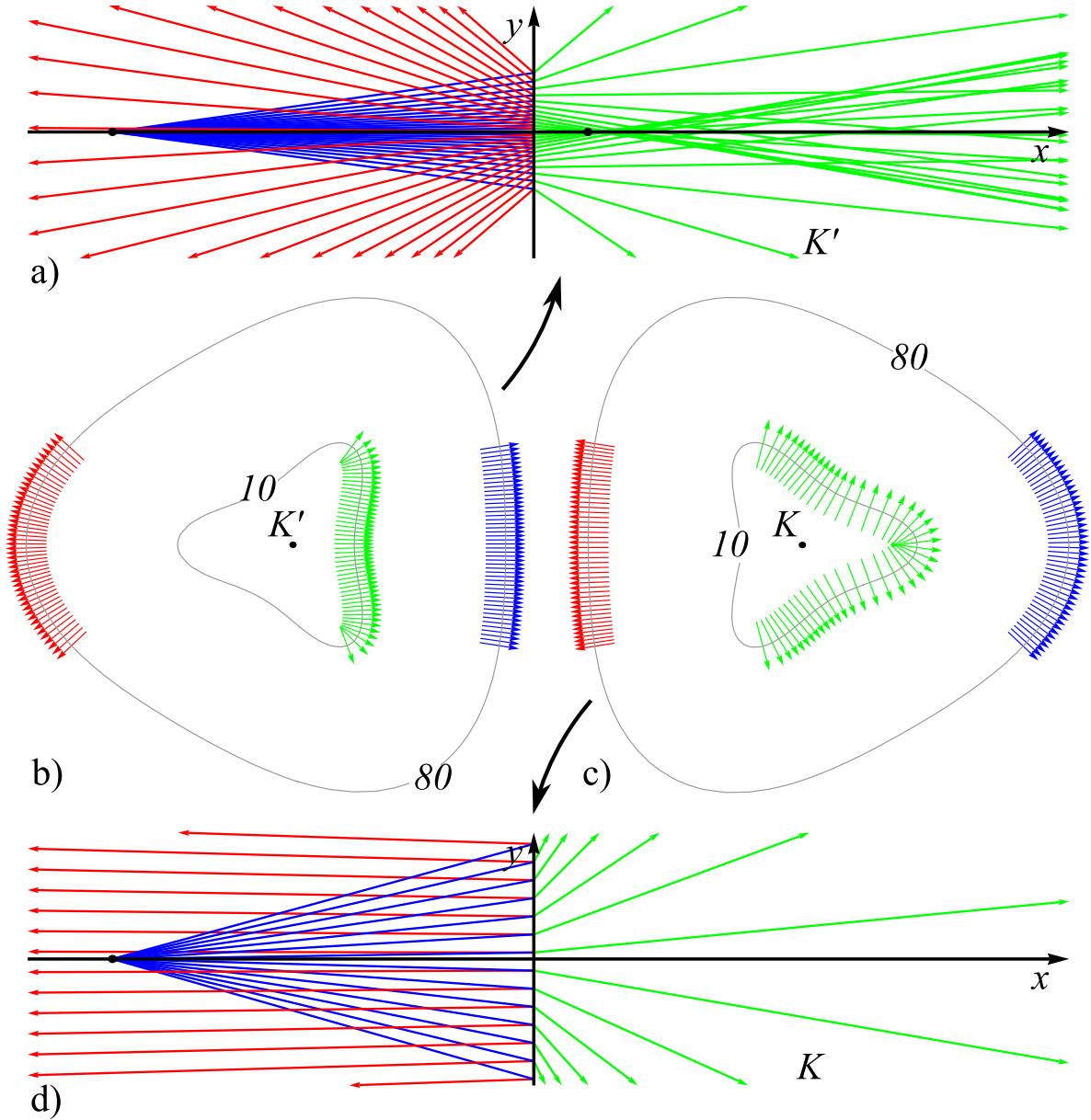}\caption{\label{fig:setup_real_and_k_space}
Real space trajectories and $k$-space directions for both valleys with barrier parameters same as in
Fig.~\ref{fig:setup_real_and_k_space-single-ray}. In figures b) and c),
blue arrows show the direction of the group velocity of incident particles
with wavevectors corresponding to the arrows' position on the $E(\mathbf{k})$
energy contour (grey solid curves). Green arrows are for the transmitted,
and red ones for the reflected electrons in both valleys. The arrows
in a) and d) show the real space trajectories of particles emitted
from the source (blue) and partially transmitted trough (green) or reflected by the junction (red). Note
that particles in valley $K'$ focus while particles in the valley
$K$ diverge. We only show trajectories that have a finite probability
of transmission at $80$ and $10\mbox{ meV}$.}
\end{figure}

\begin{figure}
\igr[width=1\columnwidth]{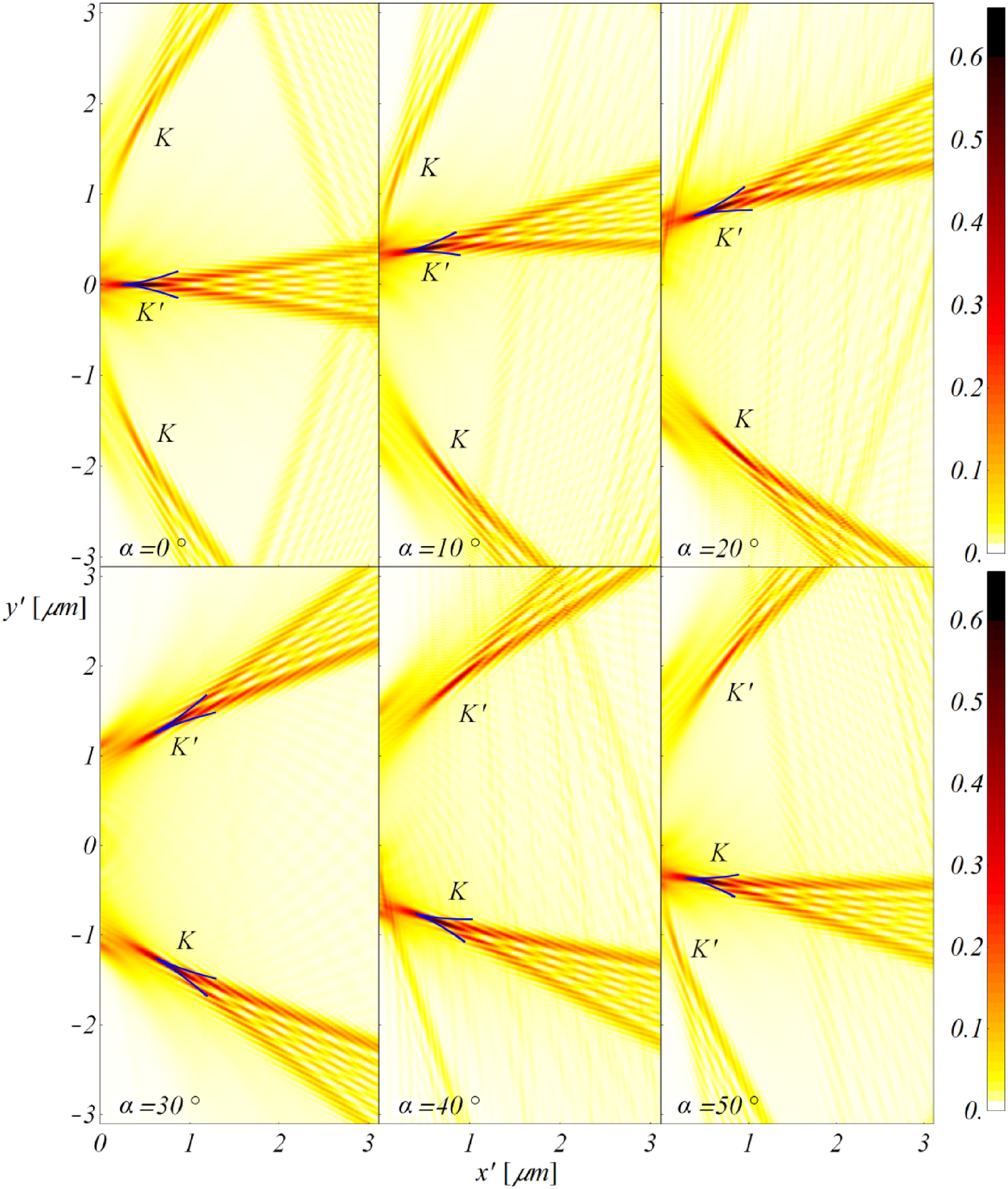}\caption{\label{fig:omega-dep-wavefunct}Total particle densities and theoretical
curves of the caustics (blue lines) for different values of $\alpha$.
Note that a given focus point only gathers particles from a single
valley. The position of the point-source is at $-2.07\mbox{ \ensuremath{\mu}m}$
on the $x'$-axis. The electrons' energy is $80\mbox{ meV}$ on the left
and $10\mbox{ meV}$ on the right side of the junction. The equation
of the sheared cusp caustics can be derived from equations~(\ref{eq:single-ray})
and~(\ref{eq:caustics})~\cite{Nye1984,Hassler2010}. }
\end{figure}

Our setup also has the capability of separating electrons in different
valleys, without the need for an unrealistic potential barrier height
of around $1\mbox{ eV}$, as in previous graphene based electron optical
setups~\cite{Garcia-Pomar2008,Wang2010}. This is demonstrated in Fig.~\ref{fig:setup_real_and_k_space},
where we follow the classical trajectories of several particles emitted
from a point source, in real and in $k$-space for the case of undistorted
samples, with the interface aligned on the $y$-axis \emph{i.e. }$w=0\mbox{ meV}$
and $\alpha=0$. It is clear from the figure that electrons from the
two valleys behave differently. Particles in valley $K$ disperse
upon passing the barrier, while those in valley $K'$ gather in a
focus point on the other side. If the barrier is not aligned with
the crystallographic orientation of the lattice, then in general there
will be two or three foci each gathering particles from different
valleys as shown in Fig.~\ref{fig:omega-dep-wavefunct}.

One key ingredient that is not captured by the ray
tracing analysis is the transmission probability of particles reaching
the junction. This can be calculated by evaluating first the expectation
values of the current operator along the $x'$-axis. Then the transmission
probability $T$ will be the ratio of the current carried by the refracted
wave in the $x'$ direction to that of the incoming wave also projected
to $x'$. The reflection probability $R$ can be calculated on a similar
basis from the reflected wave. Note that the evanescent waves do not
carry current.
\begin{figure}
\igr[width=1\columnwidth]{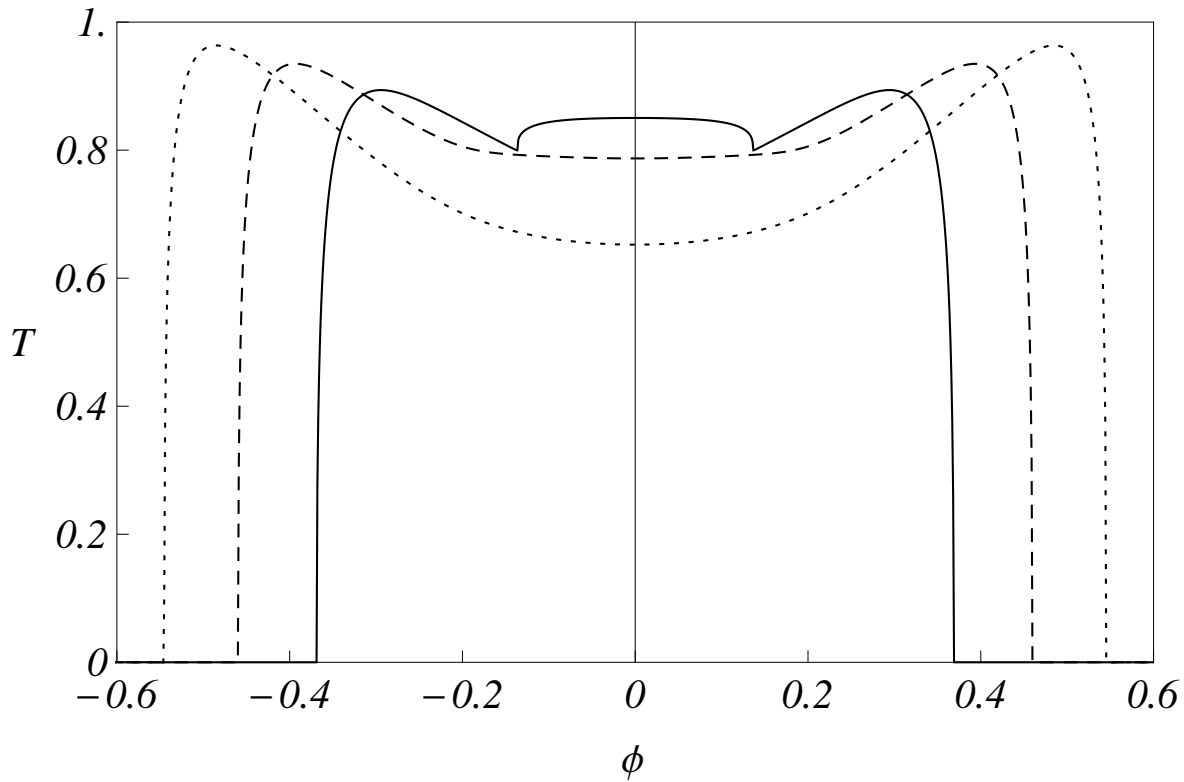}\caption{\label{fig:transmission-coeff}Quantum mechanical transmission probability
as the function of angle of incidence in valley $K$' for different
values of $w$ as in Fig.~\ref{fig:Curvatures}. The barrier parameters
are the same as in Fig.~\ref{fig:setup_real_and_k_space-single-ray}.}
\end{figure}

In Fig.~\ref{fig:transmission-coeff} we show the transmission probability
as the function the angle of incidence for different values of $w$
in the case of $\alpha=0$. Remarkably, the transmission probability
is high and roughly independent of the angle of incidence $\phi_{i}$ for particles reaching
the junction at a wide range of angles of incidence around the $x'$-axis.
This is a beneficial result of scattering
within the same band. For particles that change band during refraction
at the junction, the transmission probability is reduced, is very sensitive
to $\phi_{i}$, and vanishes exactly for perpendicular incidence~\cite{Peterfalvi2009},
unless gaps are induced by external gating~\cite{Park2011}.

For finite $\alpha$, when the junction is not aligned with the $y$-axis,
the focus point will not remain on the $x$-axis but will continuously
move with $\alpha$. This is depicted in Fig.~\ref{fig:omega-dep-wavefunct},
where the total particle density is plotted with contributions from
both valleys and for different values for $\alpha$. We also calculated
the classical curves of the caustics using~(\ref{eq:caustics}). In
the figure, these curves are represented by blue lines, the crossing
points of which approach neatly the maxima of the particle density
as expected. There are two or three focus points each gathering particles
either from the $K$, or the $K'$ valley.

\section{Effects of broken symmetry}

As shown in Ref.~\cite{Lemonik2010},
the electron-electron interaction in bilayer graphene leads to a spontaneous
breaking of the threefold rotational symmetry giving rise to a finite
value of $w$ in the effective low-energy Hamiltonian~(\ref{eq:zeHamiltonian}).
However in this description of the electron-electron interaction,
the Fermi energy as well as the energy of the excitations around it
need to be of the same order of magnitude as the Lifshitz energy \mbox{($\sim1$ meV)},
which is not the case discussed in this paper. Also, $w$ originating from the interaction can
have very different values on the two sides of the junction, which are unknown.
On the other hand, later, it was argued in Ref.~\cite{Mucha-Kruczynski2011}
that mechanical distortion leads to the same low-energy effective
theory, with a constant $w$ on a wide energy range.
A key consequence of a finite $w$ is that depending on its
value, the low energy spectrum of the system can undergo a topological
phase transition. For small magnitudes of $w$, the spectrum consists
of four Dirac cones, while at larger $w$, there are only two Dirac
cones in the spectrum. The small value of the Lifshitz energy makes
it difficult to infer directly the number of Dirac points. Therefore
it is essential to seek indirect methods to determine the value of $w$
and hence the number of Dirac points.

As presented in Fig.~\ref{fig:Curvatures}, the curvature of the
energy contours depends considerably on the value of $w$. This suggests
that the position of the focus point will be similarly sensitive.
\begin{figure}
\igr[width=1\columnwidth]{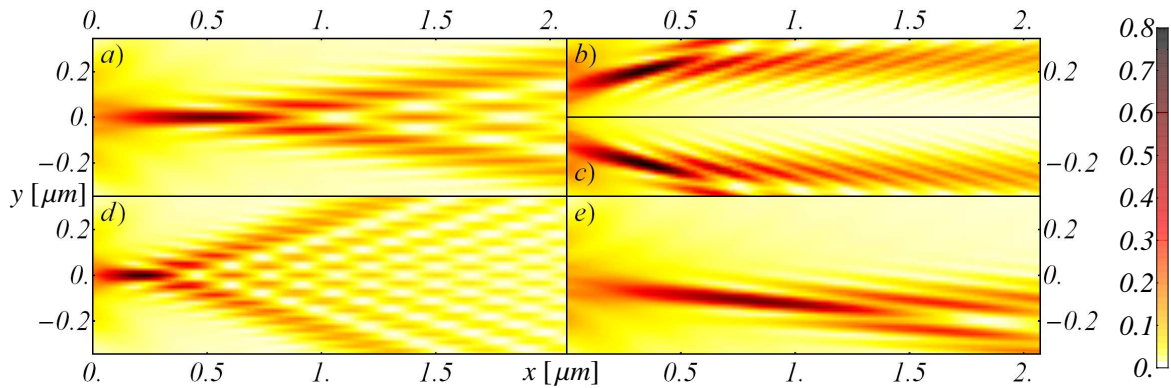}\caption{\label{fig:wfun}Calculated density of transmitted particles for different
values of $w$. The particles' energy is $80\mbox{ meV}$ on the left and
$10\mbox{ meV}$ on the right side of the junction. The source is
at $-2.07\mbox{\ensuremath{\mu}m}$ on the $x'$-axis. The sub-figures
correspond to a) $w=0\mbox{ meV}$; b) $w=6\mbox{i}\mbox{ meV}$;
c) $w=-6\mbox{i}\mbox{ meV}$; d) $w=-4\mbox{ meV}$; e) $w=2-2\mbox{i meV}$.}
\end{figure}
Indeed, this is clear from Fig.~\ref{fig:wfun} where we show particle
densities from valley $K'$ for different values of $w$. The most
important information one can extract from this data is the position
of the maxima that can be used to directly infer the value of the complex parameter $w$.
\begin{figure}
\igr[clip,width=1\columnwidth]{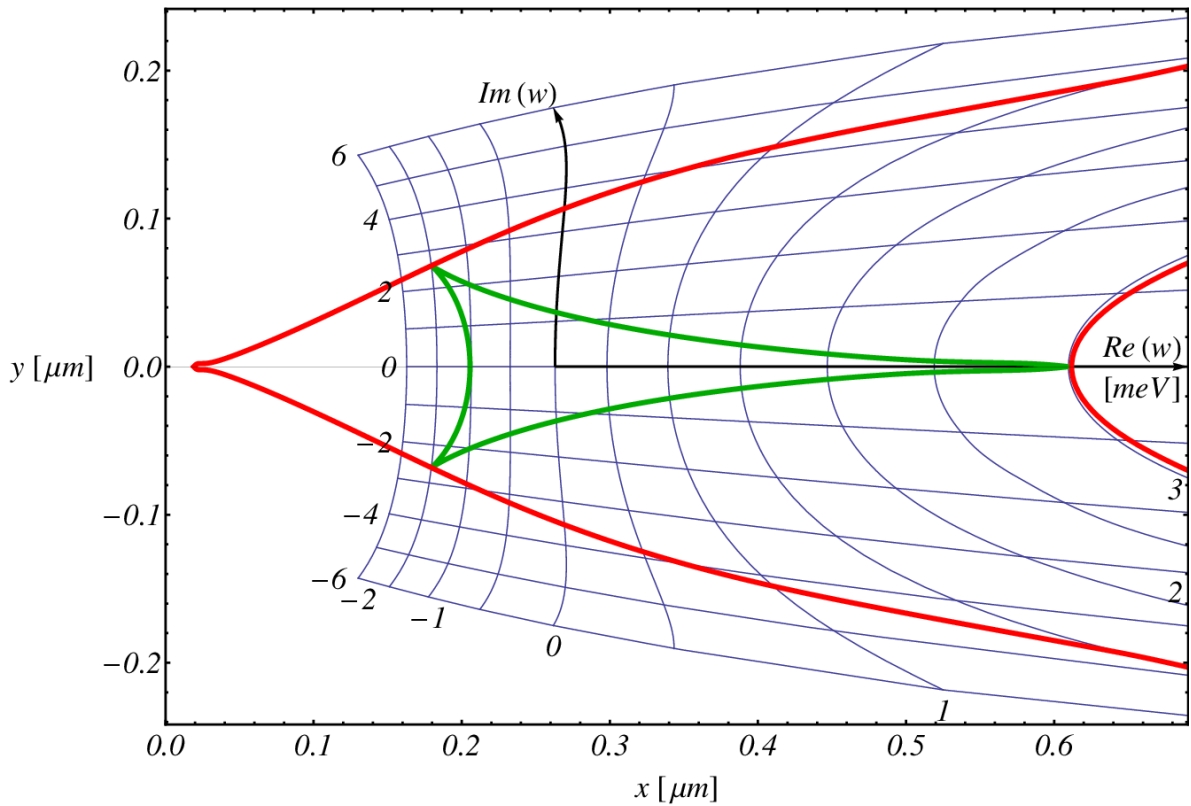}\caption{\label{fig:foc-on-grid}
A semiclassical estimation for the position
of the tip of the cusp caustic formed by electrons from the valley
$K'$ as the function of the parameter $w$ for $\alpha=0$. The electrons'
energy is $80\mbox{ meV}$ on the left and $10\mbox{ meV}$ on the
right side of the junction. The position of the source is at $-2.07\mbox{ \ensuremath{\mu}m}$
on the $x$-axis. Inside the green curve the low energy spectrum has
4 Dirac cones, between the red and green curves there are two Dirac
cones and a local minimum, while outside the red curve the spectrum
has two Dirac points but no local minimum. The nonzero value of $w$
is attributed to a distortion of the lattice that is around or less
than $1\%$ on this map~\cite{Mucha-Kruczynski2011}{}. Here we note
that the $e-e$ interaction renormalizes $w$, thus rendering it
slightly energy dependent and changing its relation to the applied
strain. This however does not affect our results qualitatively.}
\end{figure}
In Fig.~\ref{fig:foc-on-grid}, we show the semiclassical estimation
for the position of the tip of the cusp caustic as the function of
the real and imaginary part of $w$ for $\alpha=0$. From the
expression~(\ref{eq:wmech}), it is apparent that the imaginary
phase of $w$ will be $-2\theta$ reflecting a direct relation to the
geometrical deformation of the lattice. It is also worth mentioning
that if $\theta$ is a multiple of $\pi/2$, then $w$ is real, but \mbox{if $\textrm{mod}(\theta,\pi/2)=\pi/4$},
then $w$ is purely imaginary, and otherwise it is a general complex number.
Interestingly, if $\delta = \delta'$, then $w$ vanishes. This corresponds
to a 'hydrostatic' rescaling of the lattice and only affects the values of $m$ and $v_3$.
Based on the results of Ref.~\cite{Mucha-Kruczynski2011} and~\cite{mucha-kruczynski_landau_2011},
it can be show that a distortion of 1\% changes $m$ and $v_3$ just by a few percent and thus the effect
on the diffraction pattern is negligible. On the other hand, we have shown
that the position of the foci sensitively depends on the value of $w$.
Thus by experimentally investigating the focusing effect,
one can measure the external strain present in the sample.

In clean systems, the most important factor
that determines the lifetime of excitations is electron-phonon scattering.
According to recent calculations~\cite{Borysenko2010,Borysenko2011},
the mean free path for excitations in the considered energy range
around $50\mbox{ meV}$ is of the order of microns, which is
accessible by current experimental techniques~\cite{zhang_giant_2008}.

\section{Conclusion}

In this paper, we investigated the anisotropic electron optics of
bilayer graphene. We demonstrated that a moderate potential step can be used
to focus electrons within the same band in a valley-selective manner
with high transmission probability. We also investigated the effects
of broken symmetry in the electronic structure due to mechanical
distortion. The presented results clearly show that the proposed device
can be used to determine symmetry breaking and extract indirect
information about the low-energy topology of the band structure of
bilayer graphene samples. These results are expected to form key ingredients
in future device design, where one can envisage using bilayer graphene
as building blocks for more complex electron-optical devices.

\ack
The authors are grateful to E.~McCann, M.~Berry, Cs.~T\H{o}ke, V.~I.~Fal'ko,
A.~Cortijo, D.~A.~Gradinar and M.~Mucha-Kruczy\'nski for fruitful discussions.
This work has been supported by the Hungarian research funds OTKA K81492,
K76010, K75529, NK72916, NNF78842, TAMOP-4.2.1/B-09/1/KMR-2010-0002,
TAMOP-4.2.1./B-09/1/KMR-2010-0003 and the EU grant NanoCTM.

\section*{References}
\bibliographystyle{unsrt}
\bibliography{bibliography_BLG}

\end{document}